\documentclass{ws-procs10x7}

\begin{document}

\title{New Results on Deeply Virtual Compton Scattering \\ at H1, ZEUS and HERMES}

\author{F. Ellinghaus \\ (On Behalf of the H1, ZEUS and HERMES Collaborations)}

\address{University of Colorado, Boulder, Colorado 80309--0390, USA \\ 
E-mail: Frank.Ellinghaus@desy.de}

\twocolumn[\maketitle\abstract{
The process of deeply virtual Compton scattering 
will be shortly introduced, 
and the latest results from measurements at the HERA $ep$--collider at DESY will be given.
In particular, the cross section has been measured with increased statistics
at the collider experiments H1 and ZEUS, while HERMES for the first time reports measurements
of the $t$--dependence of the beam--charge asymmetry on hydrogen and 
of a beam--charge asymmetry on deuterium.}]

Hard exclusive processes such as the Deeply Virtual Compton Scattering (DVCS) 
process $\gamma^* p \rightarrow \gamma p$ can be expressed
in terms of Generalized Parton Distributions (GPDs)~\cite{Dit88,Mue94}.
GPDs give a generalized description of the
partonic structure of the nucleon, where 
the ordinary Parton Distribution functions (PDFs) and the nucleon form factors turn out to be the 
kinematic limits and the moments of GPDs, respectively~\cite {Ji97}.
Besides the fact that the DVCS process
appears to provide the theoretically most direct access to GPDs, it is also 
investigated in the framework of diffractive processes. While the DVCS process is similar to
the diffractive electroproduction of vector mesons, it 
avoids the theoretical complications of needing further non--perturbative 
information due to the vector meson distribution amplitude in the final state.
Furthermore, it is also unique among the hard scattering processes in that 
DVCS amplitudes (magnitude and phase) can be determined.
This is possible through a measurement of the interference
between the DVCS and Bethe--Heitler (BH) processes, 
in which the photon is radiated from a parton in the former and
from the lepton in the latter process. 
These processes have an identical final state, 
i.e., they are indistinguishable and thus 
the photon production amplitude $\tau$
is given as the coherent sum of the amplitudes of the DVCS ($\tau_{DVCS}$) 
and BH ($\tau_{BH}$) processes. 
The cross section 
for the exclusive leptoproduction of photons 
is then given as (see Ref.~\cite{Die97} for full expression)
\begin{equation} \label {total_gamma_xsect}
\frac{d\sigma}{dx_B \, dQ^2 \, d|t| \, d\phi} \propto
 \left| \tau_{BH} \right|^2 + 
\left| \tau_{DVCS} \right|^2 + I,
\end{equation}
where $x_B$ represents the Bjorken scaling variable, 
$-Q^2$ the virtual--photon four--momentum squared
and $t$ the square of the four-momentum 
transfer to the target.
The azimuthal angle $\phi$ is defined as the angle between the lepton
scattering plane
and the photon production plane.
At leading twist, the BH--DVCS interference term $I$ can be written as
\begin{equation} \label {I}
I =  \pm \, [c_1^I \, \cos \! \phi \, \mathrm{Re} \tilde M 
- P_l \, s_1^I \, \sin \! \phi \, \mathrm{Im} \tilde M 
], 
\end{equation}  
where +(-) denotes a negatively (positively) charged lepton beam
with longitudinal polarization $P_l$.
The DVCS amplitude 
$\tilde M$
is given by a linear combination 
of the 
nucleon form factors
with the so--called 
Compton form factors~\cite {Bel02a}, which 
are convolutions of the twist--2 GPDs 
with the hard scattering amplitude.

In the following, the H1 and ZEUS cross section measurements,
which access the squared DVCS amplitude $\left| \tau_{DVCS} \right|^2$, 
and the HERMES azimuthal asymmetry measurements,
which access the DVCS amplitude directly via the interference term $I$,  
will be presented.

\section{Measurements of Cross Sections at H1 and ZEUS}
The collider experiments H1 and ZEUS have a similar experimental setup and
thus employ a similar method for their analysis.
Since the scattered proton escapes 
detection, 
the event selection is based on the detection of two 
electromagnetic clusters and at most one reconstructed track. 
With the incoming proton defining the forward direction, 
the candidate events are subdivided into two samples
where the positron candidate is either in the forward/central ``large--$Q^2$'' region, 
or in the rear ``small--$Q^2$'' region.
A simulation of the BH contribution that describes the BH--dominated former sample can then
be used to subtract the BH contribution from the 
latter sample, which contains BH events as well as 
DVCS events plus additional background. 
At leading twist, this directly leads to the DVCS cross section since 
the $\phi$--dependent terms in the 
interference term (see Eq.~\ref{I}) vanish as the measurement 
is integrating over the azimuthal angle.

\begin{figure}
 \includegraphics[width=\columnwidth]{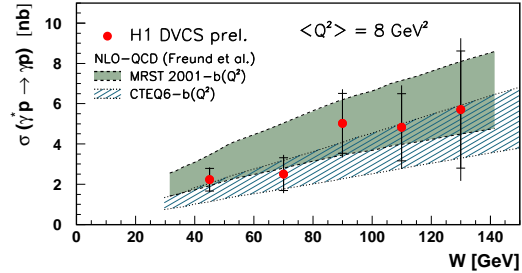}
   \caption {Preliminary H1 results on the $\gamma^* p \rightarrow \gamma p$ cross section 
     as a function of $W$ for $\langle Q^2 \rangle = 8$~GeV$^2$.
   The full error bars include the systematic error added in quadrature to the statistical error.
   The data are compared to a theoretical prediction~\protect\cite {Fre02,Fre03a} based on GPDs.
    The uncertainty in the predictions (shaded bands) is due to the unknown
    $t$--slope which is assumed to be between $5$~GeV$^{-2}$ (upper bound) and 9~GeV$^{-2}$ (lower bound).}
  \label{h1_xsec}
\end{figure}

The latest results on the $\gamma^* p$ cross section for the DVCS process, as obtained
by H1, are shown in Fig.~\ref {h1_xsec} as a function
of the  photon--nucleon 
invariant mass $W$ 
for $-t < 1$~GeV$^2$.
These exceed the published results~\cite {Adl01} by almost a factor of 4 in statistics.
The NLO QCD calculation in Fig.~\ref {h1_xsec} is based on a GPD parametrization~\cite {Fre02,Fre03a}.
Since GPDs reduce to
ordinary PDFs in a certain kinematic limit, as mentioned above,
the PDFs according to MRST2001 and CTEQ6 are 
in turn chosen to serve as an input for the model calculations.
The data agree with both models within the theoretical uncertainties, 
which are due to the up--to--now unmeasured $t$--dependence of the cross section.
Since theoretical predictions are absent as well, an exponential ansatz $e^{-b|t|}$ 
is assumed with $5$~GeV$^{-2} <~b_0 < 9$~GeV$^{-2}$ 
and $b=b_0(1-0.15 \log (Q^2/2))$ GeV$^{-2}$.
The range for the $t$-slope covers the measured range for light vector meson production,
i.e., it was chosen under the assumption that 
the production of real photons and light vector mesons have a similar $t$--slope.
Clearly, the foreseen direct measurement of the $t$--dependence will be extremely 
beneficial.

\begin{figure}
  \includegraphics[width=\columnwidth, height =7.0cm]{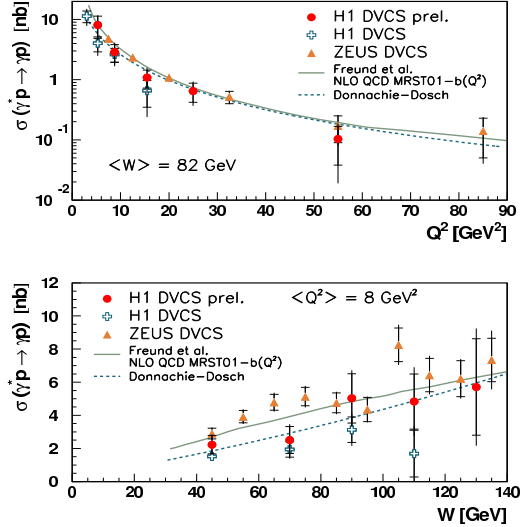}
    \caption{Preliminary H1 result together with the published results 
           from H1~\protect\cite {Adl01} and ZEUS~\protect\cite {ZEUS03} on the $\gamma^* p \rightarrow \gamma p$ 
    cross section as a function of $Q^2$ and $W$.
      The data are compared to theoretical predictions based on a GPD~\protect\cite {Fre02,Fre03a}
and on a color dipole model~\protect\cite {Don01},
using a $t$--slope of $7$~GeV$^{-2}$.}
  \label{all_xsec}
\end{figure}

Fig.~\ref {all_xsec} shows the preliminary H1 data as a function of
$Q^2$ and $W$ together with one of the results from the GPD based model calculation,
but this time at a fixed value of  $b_0 = 7$~GeV$^{-2}$. In addition, 
the already published results from H1~\cite {Adl01} and ZEUS~\cite {ZEUS03} and the results from a
color dipole model calculation~\cite {Don01} are shown.
Based on $p p$ and $\gamma^* p$ cross sections, the color dipole model predicts
the $\gamma^* p \rightarrow \gamma p$ amplitude only for forward scattering, that is at
$t = t_{min}$.
Thus, in order to calculate the integrated cross section, 
an exponential $t$--dependence with a fixed 
slope of 7~GeV$^{-2}$ is assumed as well.
The differences between the results of the GPD based model and the color dipole
model become apparent in the comparison of the cross section as a function of $W$,
where the data show a possible 
discrepancy
between the H1 and ZEUS results.
However, a direct measurement of the $t$--dependence is needed before the data 
can become conclusive.

\section{Measurements of Azimuthal Asymmetries at HERMES}
In contrast to the squared BH and DVCS amplitudes, the interference term $I$ (see Eq.~\ref{I})
does depend on the sign of the beam charge.
Therefore a  measurement of a cross section asymmetry with respect to the beam charge can isolate
the real part of the interference term, 
while the imaginary part can be isolated with a 
polarized lepton beam ($P_l \ne 0$). 
Measurements of 
azimuthal asymmetries with respect to the beam 
spin, accessing the imaginary part 
of $\tilde M$ via a $\sin \phi$ modulation, have been 
reported on hydrogen~\cite {Air01,Ste01} as well as on deuterium and neon~\cite{Ell02c}.
The extraction 
of an azimuthal asymmetry with respect to the beam 
charge, accessing the real part
of $\tilde M$ via a $\cos \phi$ modulation,
is described in the following using hydrogen and deuterium targets.

At the fixed--target experiment HERMES, events were selected 
if they contained exactly one photon and one charged track 
identified as the scattered beam lepton.
The important kinematic requirements imposed 
on the scattered lepton were $Q^2 >$ 1~GeV$^2$ and 
$W >$ 3 GeV.
The photon was identified by detecting an energy deposition in the preshower scintillator
and in the calorimeter without an associated charged track.
The polar angle $\theta_{\gamma^* \gamma}$ between the virtual and the real photon is  
required to be between 5 and 45~mrad.
Since the recoiling proton was not detected, 
events were selected if the 
missing mass $M_x$ of the reaction $e p \rightarrow e \gamma X$ corresponds to the proton
mass.
Due to the finite energy resolution the 
exclusive region is defined as $-1.5 < M_x < 1.7$~GeV based on signal--to--background studies
using a Monte--Carlo simulation.

\begin{figure}[t]
 \includegraphics[width=\columnwidth, height = 6.2 cm]{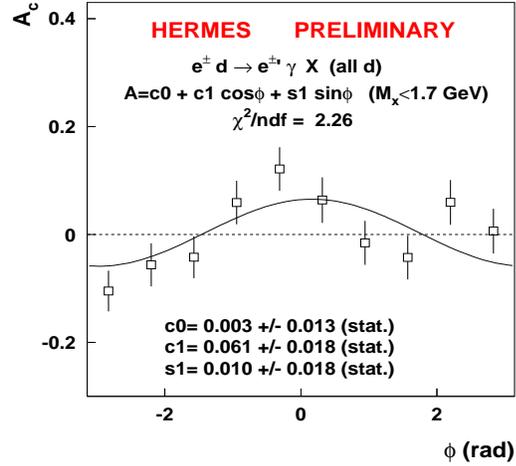}
  \caption{\label{bca_3p} Beam--charge asymmetry $A_C$ for the hard electroproduction
   of photons off the deuteron as a function of the azimuthal angle $\phi$ for the
    exclusive sample.}
\end{figure}

\begin{figure}[t]
 \includegraphics[width=\columnwidth,height = 6.1 cm]{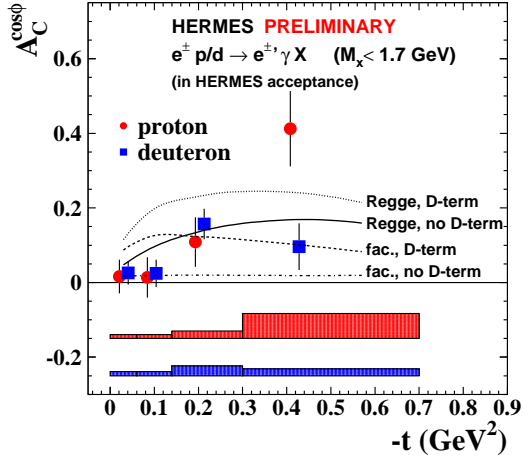}
  \caption{The $\cos \phi$ amplitude of the beam--charge asymmetry on hydrogen 
   and on unpolarized or spin--averaged polarized deuterium 
    as a function of $-t$ for the exclusive sample.
     The GPD model calculations use a factorized 
 or a Regge--inspired $t$--dependence with
    or without a D-term contribution.}
 \label{vgg_t_cos}
\end{figure}

The beam--charge asymmetry as a function of $\phi$ is calculated as
\begin{equation} \label{xsec_bca_equation}
A_C(\phi) = \frac{ N^+(\phi) - N^-(\phi)}
{N^+(\phi) + N^-(\phi)},
\end{equation}
where $N^+$ and $N^-$ represent the single photon yields normalized to 
the number of detected DIS events 
using the positron and electron beam, respectively.
It is shown in
Fig.~\ref{bca_3p} for the exclusive sample 
collected on an unpolarized or spin--averaged polarized deuterium target.
The indicated fit to the asymmetry describes the data quite well and indeed 
yields the expected $\cos \phi$ behavior.
Fig.~\ref {vgg_t_cos} shows the
$\cos \phi$ amplitudes on deuterium and hydrogen as a function of $-t$, 
which are derived from the fit to the beam--charge asymmetry in each $-t$ bin.
For both targets, the signal only becomes sizeable for larger values of $-t$.
While their slightly different behavior at large $-t$ values can be due to incoherent
scattering on the neutron in the deuteron, effects from coherent scattering on the deuteron
can be expected in the first $-t$ bin but are not apparent there.
Note that an intermediate result of the hydrogen analysis~\cite {Ell02b} 
with a preliminary $t$--averaged value of $0.11 \, \pm \, 0.04$~(stat.) $\pm \, 0.03$~(sys.)
was derived at a mean $-t$ value of 0.27~GeV$^2$, it
thus can be approximately compared to the result in the third $-t$ bin.

The theoretical calculations shown in Fig.~\ref {vgg_t_cos} are carried 
out at the average kinematics of every $-t$ bin.
They are  based on a GPD model developed in 
Refs.~\cite {Van99,Goe01}. 
The data appear to favor the model with the non--factorized $t$-dependence and a
vanishing contribution from the so--called D--term.  
It is apparent that measurements of the beam--charge asymmetry have a large
predictive power given the fact that the electron sample used was quite small.

This work was supported in part by the US Department of Energy.


\begin{thebibliography}{99}

\addcontentsline{toc}{chapter}{\numberline{}Bibliography}

\bibitem{Dit88}
F.M. Dittes et al., 
Phys. Lett. {\bf B209} (1988) 325.

\bibitem{Mue94}
D. M\"uller et al., 
Fortschr. Phys. {\bf 42} (1994) 101.

\bibitem{Ji97}
X. Ji, Phys. Rev. Lett. {\bf 78} (1997) 610,  \\
X. Ji, Phys. Rev. {\bf D55} (1997) 7114.

\bibitem{Die97}
M. Diehl et al.,
Phys. Lett. {\bf B411} (1997) 193.

\bibitem{Bel02a}
A.V. Belitsky, D. M\"uller and A. Kirchner, 
Nucl. Phys. {\bf B629} (2002) 323.

\bibitem{Fre02}
A. Freund and M. McDermott, 
Eur. Phys. J. {\bf C23} (2002) 651

\bibitem{Fre03a}
A. Freund, M. McDermott and M. Strikman, 
Phys. Rev. {\bf D67} (2003) 036001

\bibitem{Adl01} 
H1 Coll., C. Adloff et al., 
Phys. Lett. {\bf B517} (2001) 47

\bibitem{ZEUS03}
ZEUS Coll., S. Chekanov et al.,
Phys. Lett. {\bf B573} (2003) 46


\bibitem{Don01}
A. Donnachie and H.G. Dosch, 
Phys. Lett. {\bf B502} (2001) 74


\bibitem{Air01} 
HERMES Coll., A. Airapetian et al.,
Phys. Rev. Lett. {\bf 87} (2001) 182001.

\bibitem{Ste01}
CLAS Coll., S. Stepanyan et al.,
Phys. Rev. Lett. {\bf 87} (2001) 182002.

\bibitem{Ell02c}
F. Ellinghaus, R. Shanidze and J. Volmer (for the HERMES Coll.), \\ hep-ex/0212019 


\bibitem{Ell02b}
F. Ellinghaus (for the HERMES Coll.),
Nucl. Phys. {\bf A711} (2002) 171.


\bibitem{Van99}
M. Vanderhaeghen, P.A.M. Guichon and M. Guidal, 
Phys. Rev. {\bf D60} (1999) 094017.

\bibitem{Goe01}
K. Goeke, M.V. Polyakov and M. Vanderhaeghen,
Prog. Part. Nucl. Phys. {\bf 47} (2001) 401.


\end{thebibliography}
\end{document}